\documentclass[twocolumn, aps, prl,
showkeys,showpacs,nofootinbib]{revtex4-1}

\usepackage{amsmath}
\usepackage{amssymb}
\usepackage{graphicx}
\usepackage{hyperref}

\begin{document}

\title{Dual Shapiro steps of a phase-slip junction in the presence of a parasitic capacitance}

\author{Lisa Arndt}
\email{lisa.arndt@rwth-aachen.de}
\affiliation{JARA Institute for Quantum Information, RWTH Aachen University, 52056 Aachen, Germany}

\author{Ananda Roy}
\affiliation{JARA Institute for Quantum Information, RWTH Aachen University, 52056 Aachen, Germany}

\author{Fabian Hassler}
\affiliation{JARA Institute for Quantum Information, RWTH Aachen University, 52056 Aachen, Germany}

\date{February 2018}
\begin{abstract}
Bloch oscillations in a single Josephson junction in the phase-slip regime
relate current to frequency. They can be measured by applying a periodic drive
to a DC-biased, small Josephson junction. Phase-locking between the periodic
drive and the Bloch oscillations then gives rise to steps at constant current
in the I--V curves, also known as dual Shapiro steps. Unlike conventional Shapiro
steps, a measurement of these dual Shapiro steps is impeded by the presence of a parasitic
capacitance. This capacitance shunts the junction resulting in a
suppression of the amplitude of the Bloch oscillations. This detrimental effect of the parasitic capacitance can be remedied by an on-chip superinductance. Additionally, we introduce a large off-chip resistance to provide the necessary
dissipation. We investigate the resulting system by a set of analytical and
numerical methods. In particular, we obtain an explicit analytical expression
for the height of dual Shapiro steps as a function of the ratio of the
parasitic capacitance to the superinductance. Using this result, we provide a
quantitative estimate of the dual Shapiro step height. Our calculations reveal
that even in the presence of a parasitic capacitance, it should be possible to
observe Bloch oscillations with realistic experimental parameters.
\end{abstract}

\maketitle

An important goal in quantum metrology is the completion of the metrology
triangle between voltage, current, and frequency \cite{likharev:85,Pekola:13}.
Bloch oscillations in small Josephson junctions provide the last needed link
between current and frequency and thus, have the potential to close the
metrology triangle \cite{Averin:85}. In order to observe these oscillations,
a large impedance is needed to reduce the charge fluctuations and reach the
Coulomb blockade regime \cite{schmid:83,grabert}. In addition, the Josephson
junction needs to be operated as a
quantum phase-slip junction, the dual counterpart of the Josephson junction
\cite{mooij:06,pop:10,astafiev:12,masluk:12}.

Bloch oscillations can be measured in a quantum phase-slip junction biased with
a current $I_0$ and irradiated with microwaves of frequency $\omega_0$. When
$I_0$ is an integer multiple of $e\omega_0/\pi$, the incident radiation
phase-locks with the Bloch oscillations in the junction. This leads to dual
Shapiro steps in the I--V curve at constant current $I_0$
\cite{Averin:85,likharev:85}. The observation of Coulomb blockade is a first
prerequisite to observe dual Shapiro steps. It has already been
seen in different systems, \emph{e.g.}, nanowires
\cite{hongisto:12,Lehtinen:12}, Cooper pair transistors
\cite{haviland:94,Corlevi:06}, and single Josephson junctions
\cite{kuzmin:91,haviland:91}. However, first attempts to experimentally
demonstrate dual Shapiro steps \cite{kuzmin:91,Lehtinen:12} did not reveal
clear current steps at integer multiples of $e\omega_0/\pi$ in the I--V
curves. The experimental difficulties are connected to the detrimental effects
of unwanted parasitic capacitances due to the biasing lines. As a remedy, a
large resistance has to be placed close to the junction which leads to
excessive heating, washing out the dual Shapiro steps \cite{kuzmin:91}. In
 Ref.~\cite{manucharyan}, it has been argued that the complementary
requirements --- large resistance for good current biasing and small resistance
for small heating --- are irreconcilable. Furthermore, it was proposed to replace the large resistance by a reactive alternative, a superinductance \cite{koch:09}, which reduces the charge fluctuations without introducing heating \cite{Corlevi:06}. The effect of superinductances on dual Shapiro steps has been investigated in Refs.~\cite{guichard:10,marco:15}, but, without taking into account parasitic capacitances. In this work, we combine an on-chip superinductance, screening the parasitic capacitance, with a large off-chip resistance, dissipating the excess energy. For this setup, we provide analytical results for the height of the dual Shapiro steps and verify our findings using numerical simulations. Our predictions are relevant for the current experimental efforts towards observing dual Shapiro steps. With the recent experimental progress towards building superinductances using Josephson junction arrays \cite{Corlevi:06,manucharyan:09, bell:12,masluk:12,altimiras:13,weissl:15} or coils \cite{Fink:16}, we are optimistic of experimental verifications of our theoretical predictions.

\begin{figure}[tb]
	\centering
	\includegraphics{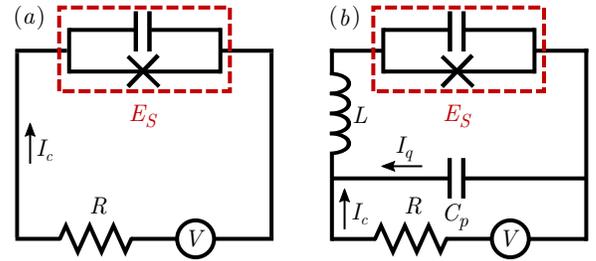}
	\caption{%
		(color online) ($a$) The ideal setup for the observation of dual Shapiro
		steps consists of a phase-slip junction $E_S$, in series with a
		resistance $R$ and an ideal voltage source $V$. The phase-slip
		junction is formed by a Josephson junction in parallel to a
		capacitance. The voltage source $V$, together with the large
		resistance $R$, acts as a current source providing a current
		$I_c$. This current can be computed by a direct measurement of the voltage drop across the resistance. ($b$) In a realistic setup, the phase-slip junction is
		shunted by an unwanted parasitic off-chip capacitance $C_p$,
		causing the part $-I_q$ of the current to flow past the
		junction. To remedy this unwanted effect, an additional
		on-chip superinductance $L$ is introduced.
	}\label{fig:setup}
\end{figure}

The article is organized as follows. We start by investigating the ideal
regime of high impedance, where the characteristic impedance, $Z=\sqrt{L/C_p}$,
formed by the superinductance $L$ and the parasitic capacitance $C_p$ is
larger than the quantum resistance $R_Q=h/4e^2$. Next, we examine the
experimentally more relevant regime $Z\simeq R_Q$, where quantum effects play
an important role. Finally, we validate our analytic results by comparing
them to numerical simulations.

The ideal setup for the observation of Bloch oscillations is shown in
Fig.~\ref{fig:setup}($a$). It is the electrical dual of the conventional Shapiro
step experiment \cite{mooij:06}. It consists of a voltage source in
series with a large resistance and a phase-slip junction. The phase-slip
junction can be realized by a (small) Josephson junction with a finite
capacitance \cite{Averin:85,likharev:85,zorin:06}. The voltage source drives an incident signal composed of a DC and an AC component at frequency $\omega_0$ and is given by $V(t) = V_0 +
V_\text{ac} \sin(\omega_0 t)$. Here, we treat the relevant case of $V_\text{ac} \ll V_0$. In an ideal situation, the
large resistance $R$ together with the voltage source constitutes an almost
ideal current source with value $V(t)/R$. However, in a realistic situation,
an unwanted parasitic capacitance $C_p$ tends to shunt the Josephson junction [Fig.~\ref{fig:setup}($b$)]. This could be either due to direct coupling between the leads to and from the junction caused by insufficient separation or due to indirect coupling between the leads through a common ground plane. As this parasitic capacitance shunts part of the current injected by the
source, the Josephson junction will not be perfectly current biased which
results in the Bloch oscillations being washed out.

As a large enough resistance $R  \gtrsim R_Q$ causes to much heating,
superinductances have been proposed as a reactive alternative in order to
suppress fluctuations in the current \cite{koch:09,manucharyan:09,masluk:12}.
However, as the system is constantly driven the energy still has to be
dissipated at some point. In Ref.~\cite{manucharyan}, the authors have
proposed to implement the driving as well as the dissipation by a microwave
transmission line. In this work, we treat an alternative setup where we combine
the idea of the superinductance to protect against parasitic capacitances in
the inner current loop with a resistance in series with the voltage
source. The key point is that the resistance does not have to be close to the
Josephson junction as the superinductance serves to protect against the
parasitic capacitance. Rather, the task of the resistance is to turn the voltage
source into a current source and to dissipate excess energy so that the
system may settle into a stationary state.

The system, we propose in order to observe Bloch oscillations, is given by
Fig.~\ref{fig:setup}($b$). In this setup, an inductance $L$ serves to protect
against the parasitic capacitance $C_p$. We denote the current in the outer loop by $I_c$. The current in this loop is driven by the voltage source and it is stabilized by the presence of a resistance $R$. Under the assumption that
$R\gg R_Q$, the dynamics of $I_c$ can be treated classically (see also below). Thus, the current $I_c$ can be measured via the voltage drop over the resistance $R$. On the other hand, the current in the inner loop, denoted by $I_q$, flows without dissipation and is treated as a quantum mechanical operator.
The current fluctuations are suppressed by the characteristic impedance $Z=
\sqrt{L/C_p}$. We need the
Josephson junction in the transmon regime with $E_J \gtrsim E_C$ \cite{houck:09}.
In this case, the ground state energy is approximately given by \cite{koch:07}
\begin{equation}\label{eq:ground_state}
	E_S(q) \simeq  E_C^{1/4} E_J^{3/4} e^{-(8 E_J/E_C)^{1/2}} \cos(\pi q/e)
\end{equation}
with $q$ the charge that is accumulated on the capacitor plate. Equivalently,
the voltage across the Josephson junction is given by $V_S(q) = V_c \sin(\pi
q/e)$ with $e V_c/\pi \simeq E_C^{1/4} E_J^{3/4} e^{-(8 E_J/E_C)^{1/2}} $. The
capacitance associated with this phase-slip junction is given by
$C_S=e/\pi V_c$. In order for Eq.~\eqref{eq:ground_state} to be a good
approximation of the energy stored in the phase-slip junction, the driven
system has to stay in the ground state with vanishing Landau-Zener processes
\cite{koch:07}. This restricts the drive frequency to $\omega_0 \ll
E_J^2/\hbar E_C$, which we take to be true throughout this work. 

The step to a quantum description of the problem is performed by
introducing the loop charge operators $\hat Q_{c[q]} =
\int_{-\infty}^t\!dt'\,\hat I_{c[q]}(t')$ that denote the charge that has
flown in the classical [quantum] loop up to a time $t$ \cite{ulrich:16}.
Kirchhoff's voltage law then demands that
\begin{align}
	R \dot{\hat Q}_c &= V(t) \!+\!\frac{\hat{Q}_q}{C_p} , \label{eq:kirchhoff_2}\\ L 
	(\ddot{\hat Q}_c & \!+\! \ddot{\hat Q}_q) \!+\!
	V_S\Bigl(\hat Q_c \! +\! 
		\hat Q_q\Bigr) \!+\! \frac{\hat{Q}_q}{C_p} =0.\label{eq:kirchhoff_1}
\end{align}
Here, we have included only the noiseless, classical part of the voltage source since we assume that the system is operated at a low enough temperature $T$, with $k_B T \ll eV_c$. In this case, the thermal noise of the resistance is negligible (for analysis including thermal noise, see~\cite{suppl}).

For $R \gg R_Q$, the quantum fluctuations of $\hat Q_c$ are suppressed far below $2e$ \cite{grabert}. As a result, 
the operator can be simply replaced by its quantum-mechanical expectation
value $Q_c = \langle \hat Q_c \rangle$. The motion of $\hat{Q}_q$ is given by Eq.~\eqref{eq:kirchhoff_1}. This equation describes a non-dissipative dynamics of $\hat{Q}_q$ and thus, can be described by a Schr\"odinger equation. The explicit form of the Hamiltonian that leads to the equation of motion for $\hat{Q}_q$ is given by
\begin{equation}\label{eq:Ham_Qq}
	\hat{H}=\frac{ \hat{\Phi}_q^2}{2L}+\frac{\hat{Q}_q^2}{2C_p}+L \hat{Q}_q
	\ddot{Q}_c-\frac{e V_c}{\pi}\cos[\pi(\hat{Q}_q+Q_c)/e],
\end{equation}
where $\hat{\Phi}_q$ is the canonically conjugate variable of $\hat{Q}_q$ with
$[\hat{Q}_q,\hat{\Phi}_q]=i\hbar$ \cite{commut}. Note that this Hamiltonian is time-dependent, where the time-dependance is parametrized by $Q_c$. Thus, the problem reduces to finding the
solution of the time-dependent Schr\"odinger equation $i \hbar \partial_t
\psi(Q_q; Q_c, t) = \hat H \psi(Q_q; Q_c, t)$ for the quantum loop charge
coupled to the equation of motion
\begin{equation}\label{eq:Motion_Qc}
	R \dot{Q}_c =\langle \psi|\hat{Q}_q| \psi\rangle /C_p + V(t)
\end{equation}
for the classical loop charge. The dynamics of the circuit are governed by
three distinct rates: the plasma frequency of the quantum charge
$\omega_q=1/\sqrt{LC_p}$, the plasma frequency of the classical charge
$\omega_c=1/\sqrt{LC_S}$, and the RC-rate $\omega_{R}=1/R C_S $ with which the
motion is damped. The ratio $\omega_c/\omega_R$ is the quality factor of the
classical charge dynamics and measures the relative importance of the first- with respect to
the second-order time derivative of $Q_c$. For $\omega_R > \omega_c$, the
system can show hysteretic behavior, which makes it very unfavorable for the
accurate observation of dual Shapiro steps. In the following, we will therefore
focus on the overdamped regime with $\omega_c \gg \omega_R$.

In order to analyze the problem, we first treat the case of a large
characteristic impedance $Z \gg R_Q$. In this case, the charge $\hat Q_q$ as
well as its fluctuations in the parasitic loop remain small compared to $2e$. This allows us to linearize Eq.\eqref{eq:kirchhoff_1} and treat $\hat Q_q$ classically. Next, we insert the obtained solution for $\hat Q_q$ into the equation for the
classical loop charge and neglect the second order derivatives, since we are
interested in the overdamped regime. This leads to 
\begin{equation}\label{eq:lin}
R \dot Q_c +  \frac{V_c\sin(\pi Q_c/e)}{1+ (C_p/C_S) \cos(\pi Q_c/e) } =
V(t).
\end{equation}
For vanishing $C_p$, this reduces to the (dual of the) RSJ-model of the
conventional Shapiro steps in the overdamped regime \cite{likharev}. The
resulting voltage steps at fixed DC current $I_0$ through the outer loop are dual to
the conventional Shapiro steps. The position $I_0$ originates from the phase-locking of the external drive frequency $\omega_0$ to the frequency of the
Bloch oscillation $I_0\pi/e$. We find that the first step appears at
$I_0 = e \omega_0/\pi$ centered around the value $V_0= V_c
\sqrt{1+(\omega_0/\omega_R)^2 } [1 +  (3/8)(\omega_R/\omega_0)^2 (C_p/C_S)^2]
$ of the DC-voltage. The step height is given by
\begin{equation}\label{eq:height}
	\Delta V = \frac{V_\text{ac} [ 1 +  (1/4)(C_p/C_S)^2 ]}{\sqrt{1+(\omega_0/\omega_R)^2 } }
\end{equation}
to leading order in $\omega_R/\omega_0$, $C_p/C_S$, and $V_\text{ac}/V_c$ \cite{suppl}. The
temperature has a negligible effect on the step as long as $k_B T \ll e \Delta
V$ \cite{likharev,suppl}. Note that in this regime the presence of the
parasitic capacitance increases the height of the dual Shapiro step.

\begin{figure}[tb]
	\centering
	\includegraphics{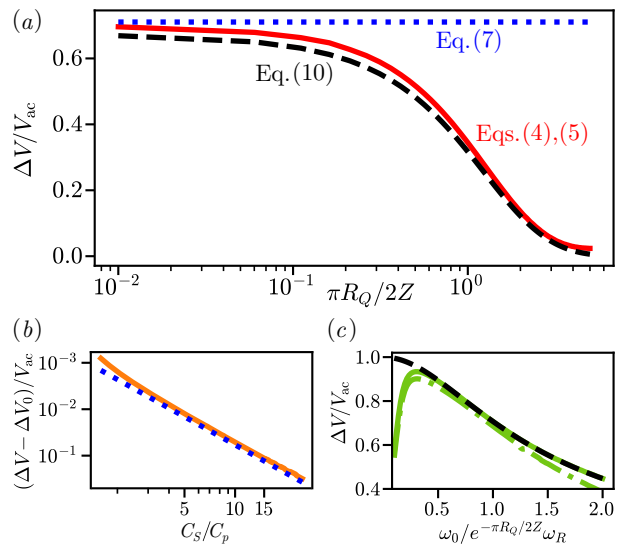}
	\caption{%
		(color online) ($a$) This panel shows the height of the first dual Shapiro step
		as a function of the characteristic impedance $Z$ for
		$\omega_c/\omega_R=2$, $\omega_q/\omega_R=3$,
		$\omega_0/\omega_R=1.1$, and $V_\mathrm{ac}=0.1 V_c$. The
		solid, red line shows the result obtained by a numerical
		calculation of the coupled system of Eqs.~\eqref{eq:Ham_Qq}
		and \eqref{eq:Motion_Qc}. The dotted, blue
		line shows the analytic approximation of the step height in the
		high impedance regime $Z\gg R_Q$ in Eq.~\eqref{eq:height}. The dashed, black line shows the analytical result to the ground state approximation in Eq.~\eqref{eq:Step_Likarev}. The ground state approximation provides a good, conservative
		approximation for the results of the quantum simulation,
		especially for $Z\ll R_Q$. ($b$) Comparison of the analytic approximation
		of Eq.~\eqref{eq:height} to the numerical solution of
		Eq.~\eqref{eq:lin} (solid, orange line). The reference step height $\Delta V_0$ corresponds to $C_p=0$. As expected the
		approximation works best at large $C_S/C_p$.
		($c$) Comparison of the analytic
		result of the ground state approximation in the overdamped
		regime [Eq.~\eqref{eq:Step_Likarev}] to a numerical solution
		of the classical equation of motion
		[Eq.~\eqref{eq:Motion_Qc_Ground}]. The solid, green line shows the
		numerical result for $\omega_R e^{-\pi R_Q/4Z}/\omega_c=0.01$,
		the dash-dotted line for $\omega_R e^{-\pi R_Q/4Z}/\omega_c=0.5$. The analytical approximation fits the numerical data well,
		especially for the strongly overdamped regime (solid line).
	}\label{fig:steps_comp}
\end{figure}

Experimentally more relevant is the regime $Z \simeq R_Q$. In order to obtain
analytical results in this regime, we assume the plasma frequency of the
quantum charge to be large. In particular, we demand that the parasitic
capacitance is small enough such that the relations $\omega_q \gg \omega_0$
and $ \omega_q \gg e V_c /\hbar \simeq \omega_R R/R_Q$ hold. Under these conditions,
the quantum loop charge stays in the ground state of $\hat H$ during
the course of the evolution, given the system is initially at sufficiently low
temperatures with $k_BT \ll \hbar \omega_q$. Specifically, as $\omega_q \gg
\omega_0$ and $\hbar \omega_q \gg e V_c$, we can neglect the last two terms in
\eqref{eq:Ham_Qq}. Then, the ground state wave function is that of a harmonic oscillator and is given by
\begin{align}\label{eq:gs_wf}
	\psi_0(Q_q,Q_c,t) =\frac{Z^{1/4}}{\pi^{1/4} \hbar^{1/4}}
e^{- (Z/2 \hbar) Q_q^2 - i \omega_q t/2  }.
\end{align}
Using this wave function in order to calculate the expectation value of
$\hat{Q}_q/C_p$, the equation of motion of the classical charge
reduces to \cite{suppl}
\begin{equation}\label{eq:Motion_Qc_Ground}
	L \ddot{Q}_c+R \dot{Q}_c + e^{-\pi R_Q/2Z} V_c  \sin (\pi Q_c/e)=V(t),
\end{equation}
valid to lowest order in $e V_c/\hbar \omega_q$ and $\omega_0/\omega_q$. Note that in this equation the sole effect of the parasitic capacitance is to
reduce the critical voltage of the phase-slip junction $V_c$ by a factor
$e^{-\pi R_Q/2Z}$. This is the main result of our paper. It implies that the
effect of the parasitic capacitance is shielded by the inductance as long as
the characteristic impedance $Z$ is larger than $\pi R_Q/2 \approx
10\,$k$\Omega$. Indeed, we find that in the overdamped regime $\omega_c \gg
e^{-\pi R_Q/4Z}\omega_R$, where we can neglect the second-order time
derivative, we can again calculate the step height analytically \cite{likharev}.
Assuming that $V_0>e^{-\pi R_Q/2Z}V_c$ the first step at the current $I_0 = e
\omega_0/\pi$ appears at the voltage $V_0 = V_c \bigl[e^{-\pi R_Q/Z} +
(\omega_0/\omega_R)^2\bigr]^{1/2}$. The height (in voltage) of the step at
constant current is on the other hand given by
\begin{align}\label{eq:Step_Likarev}
	&\Delta V = \frac{V_\text{ac}}{\sqrt{1+e^{\pi
	R_Q/Z}\,(\omega_0/\omega_R)^2 }},
\end{align}
valid to first order in $V_\mathrm{ac}/e^{-\pi R_Q/2Z}V_c $.

In order to confirm these results, we have performed numerical calculations.
First, we solved the coupled system between the Hamiltonian in Eq.~\eqref{eq:Ham_Qq} and the equation of motion in Eq.~\eqref{eq:Motion_Qc} numerically. This was done by calculating the time dependent Schr\"odinger equation in the basis of the harmonic oscillator using the Crank-Nicolson method, while the classical equation of motion was solved using the backward Euler method. In Fig.~\ref{fig:steps_comp}($a$), we compare our numerical results to our analytical solutions from Eq.~\eqref{eq:height} and Eq.~\eqref{eq:Step_Likarev}. We find that the approximation in the high impedance regime provides an upper limit to the quantum simulation since the calculation is only valid for $Z\gg R_Q$. The ground state approximation gives a good, conservative approximation for the size of the step in the quantum simulation, especially for $Z\ll R_Q$. This is due to the fact, that the condition for the ground state approximation $\omega_q\gg e V_c/\hbar \simeq \omega_c \sqrt{Z/R_Q}$ is better fulfilled in this regime. 

Second, we solved Eq.~\eqref{eq:lin} numerically using the
backward Euler method. The result can be found in Fig.~\ref{fig:steps_comp}($b$) together with our analytical result from Eq.~\eqref{eq:height}. As expected, the analytic result
works best at small $C_p/C_S$.
Third, we solved Eq.~\eqref{eq:Motion_Qc_Ground} numerically, again using the backward Euler method. In Fig.~\ref{fig:steps_comp}($c$), we compare these
results to our analytic approximation for the overdamped regime in
Eq.~\eqref{eq:Step_Likarev} for two different values of $\omega_R e^{-\pi
R_Q/4Z}/\omega_c$. We find that for $\omega_c\gg e^{-\pi R_Q/4Z}\omega_R $ the
numerical calculation agrees very well with the numerical result, as long as
the assumption $V_0>e^{-\pi R_Q/2Z}V_c$ is valid.  

In order to better illustrate the dependence of the step size on the
parameters, we used the numerical results from the ground state approximation
in Eq.~\eqref{eq:Motion_Qc_Ground} to create a color plot, which shows the
dependence of the step size on all relevant parameters. Here, we can clearly
see that the maximum step size can only be achieved in the overdamped regime.
For $\omega_c\lesssim e^{-\pi R_Q/4Z} \omega_R $, the step sizes becomes
smaller and hysteresis begins to occur, making it very unfavorable for a
precise measurement of the step position. For $\omega_c > e^{-\pi R_Q/4Z} \omega_R $, the
results do not sensitively depend on the parameters used. The optimal step
size can be achieved for driving frequencies $\omega_0 \approx 0.5 e^{-\pi R_Q/2Z} \omega_R
$, with a value $\Delta V$ reaching over 80\% of the maximal
theoretical step size $V_\text{ac}$.

\begin{figure}[tb]
	\centering
	\includegraphics{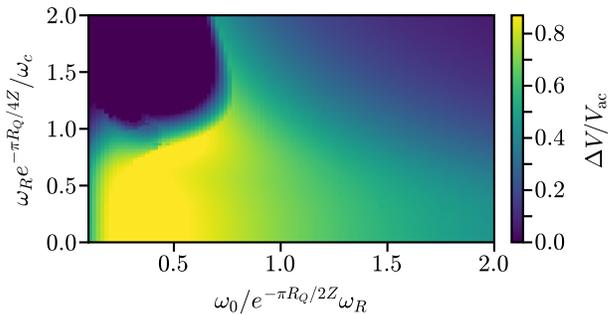}
	\caption{%
		(color online) Size of the first dual Shapiro step
		in the ground state approximation as a function of the quality factor and the drive frequency. It was obtained by solving
		Eq.~\eqref{eq:Motion_Qc_Ground} numerically for
		$V_\mathrm{ac}=0.1 e^{-\pi R_Q/2Z} V_c$. It can be seen that the
		largest steps appear in the overdamped regime $\omega_c\gg
		e^{-\pi R_Q/4Z}\omega_R $ for a drive frequency
		$\omega_0\approx 0.5 e^{-\pi R_Q/2Z}\omega_R$.
	}\label{fig:steps_ground}
\end{figure}

Next, we comment on the experimental feasibility of the observation of
dual Shapiro steps. Nowadays, it is possible to fabricate Josephson junctions
with $E_J, E_C/2\pi\hbar\simeq 10\,$GHz \cite{Corlevi:06}, which results in
critical voltages $V_c$ of the order of $10\,\mu$V ($C_S\simeq 5\,$fF). In
order to avoid Landau-Zener processes, the drive frequency $\omega_0/2\pi$ should remain well
below $10\,$GHz. Therefore, the dual Shapiro steps will appear at currents of
the order of nano-amps. In this context, note that a current standard formed
by Shapiro steps is readily parallelizable in order to achieve larger values
\cite{Pekola:13}. Modern fabrication techniques allow for on-chip inductances
of the order of $500\,$nH \cite{bell:12,masluk:12,Fink:16}. For our purposes,
we would require parasitic capacitance of the order of $100\,$fF in order to
obtain an impedance $Z=\sqrt{L/C_p}$ of about $3\,\mathrm{k}\Omega$. Given the
many groups in different fields working on the fabrication of
superinductances, we are confident that this will be reached soon. In addition, sufficiently low temperature as well as effective noise filtering is required to prevent the dual Shapiro steps from being washed out \cite{suppl}.

In conclusion, we have analyzed the dual Shapiro step height in the presence of a superinductance and a parasitic capacitance. We have described the system by a Schr\"odinger equation coupled to a classical equation of motion. In the limit $\omega_q \gg \omega_0,
eV_c/\hbar$, the quantum system remains in the ground state and only the
classical equation of motion has to be solved. We have provided an analytical
expression for the dual Shapiro step height in the overdamped limit
$\omega_c\gg  e^{-\pi R_Q/4Z} \omega_R $. The leading effect of the parasitic capacitance is a reduction of
the critical voltage of the phase-slip junction $V_c$ by a factor of $e^{-\pi
R_Q/2Z}$. Thus, the effect of the
parasitic off-chip capacitance can be remedied by an on-chip inductance, as
long as the characteristic impedance $Z$ is of the order of $\pi R_Q/2\approx
10\,\mathrm{k}\Omega$. Additionally, we have shown the
dependence of the step height on $\omega_q$ by deriving an expression for the
step height in the limit of high characteristic impedance. Finally, we have
performed numerical simulations to validate the analytical results. Throughout this work,
we have chosen to neglect the effects of thermal noise of the resistance and the influence of the stray capacitance parallel to the inductance. This is because thermal effects can be made small by working at low temperature and the stray capacitance
is usually much smaller than the parasitic capacitance. A detailed analysis of these effects is left for the future.

We acknowledge fruitful discussions with M. Devoret, J. Fink, and M. Peruzzo. A. R. acknowledges the support of the Alexander von Humboldt foundation.

\onecolumngrid\clearpage\setcounter{equation}{0}
\renewcommand\theequation{S\arabic{equation}}
\section{Supplement}\setcounter{page}{1}

\subsection{Analytical approximation at large impedance}
First, we will analyze the system in the regime of large characteristic
impedance $Z\gg R_Q$. In this case, the charge $\hat Q_q$ as
well as its fluctuations in the parasitic loop remain small compared to $2e$. This allows us to linearize Eq.\eqref{eq:kirchhoff_1} and treat $\hat Q_q$ classically. Next, we insert the obtained solution for $\hat Q_q$ into the equation for the
classical loop charge. In the overdamped regime $\omega_c\gg\omega_R$, we can then additionally neglect the second order time derivatives which leads to Eq.~\eqref{eq:lin}. If we now
expand this equation up to second order in small $C_p/C_S$ we obtain
\begin{align}\label{eq:lintwice}
\dot N /\omega_R+  \sin(N)-  (C_p/2C_S) \sin(2N)+
(C_p/2C_S)^2\left[\sin(N)+\sin(3N)\right]  =v(t), 
\end{align}
where $v(t)=V(t)/V_c$ is the normalized voltage and $N=\pi Q_c/e$ the
normalized charge in the classical loop. For vanishing $C_p$ this again
reduces to the (dual of the) RSJ-model of the conventional Shapiro steps in
the overdamped regime \cite{likharev}
\begin{equation}\label{eq:likharev}
\dot N/\omega_R +  \sin(N)  =v(t).
\end{equation}
For the RSJ-model we can calculate the step size analytically to leading order
in $V_\text{ac}/V_0$ \cite{likharev}. 

We start by calculating the solution to
Eq.~\eqref{eq:likharev} without an AC drive $v(t)=v_0$ and obtain for $v_0>1$
\begin{equation}\label{eq:solutionDC}
N_\text{DC}=2\arctan \left(\frac{i_0}{v_0-1}\tan \left[\frac{\omega_Ri_0 (t-t_0)+\frac{\pi}{2})}{2}\right]\right)-\frac{ \pi}{2},
\end{equation}
where $i_0=I_0 R/V_c=\sqrt{v_0^2-1}$ is the normalized average current running through the classical loop. The initial condition can always be fulfilled by choosing an according $t_0$.

Now, we will include a small AC drive in addition to the DC bias and expand the
solution to Eq.~\eqref{eq:likharev} in a Taylor series
$N=N_\text{DC}+v_\text{ac} N_\text{AC}$. We also need to take into account
that the AC drive can change the average voltage. Therefore, we will
formally expand the DC bias in a Taylor series $v(t)=v_0+v_\text{ac} [v_s+\sin
(\omega_0 t)]$, too. Here, $v_s$ acts as the additional influence on the
average voltage due to the AC drive. If we insert both expansions into
Eq.~\eqref{eq:likharev} we obtain a differential equation for $N_\text{AC}$
\begin{equation}\label{eq:Acdifferential}
\dot N_\text{AC}/\omega_R +  \cos(N_\text{DC})N_\text{AC} =v_s+\sin(\omega_0 t).
\end{equation}
This equation does not need to be solved in order to obtain the dual Shapiro step size.
Instead, we will specifically consider the first step where $\omega_0=\omega_R
i_0$. At the step, the average current has to remain constant $\overline{\dot
N_\text{AC}}=0$. Using Eq.~\eqref{eq:Acdifferential} this restraint can be
rewritten to obtain
\begin{equation}\label{eq:ConditionStep}
\overline{\left[v_s+\sin(\omega_Ri_0 t)\right]/\dot N_\text{DC}}=0,
\end{equation}
which only depends on the solution without an AC drive that we already
obtained in Eq.~\eqref{eq:solutionDC}. As a final result we thus obtain an
equation for $v_s$
\begin{equation}\label{eq:ConditionStept0}
v_s=-\cos(\omega_Ri_0t_0)/2v_0.
\end{equation}
We find that depending on the phase shift $t_0$ between the Bloch oscillations
and the AC drive, $v_s$ and thus the position on the voltage step varies. This results in a maximum size of the first step $\Delta
V_0=V_\text{ac}/v_0$, which coincides with Eq.~\eqref{eq:height} in the limit
$C_p\rightarrow 0$.  

Next, we want to obtain the step size for finite $C_p$ up
to second order in small $C_p/C_S$. Analog to the previous calculation, we
first need to find a solution to Eq.~\eqref{eq:lintwice} in the case without
an AC drive. We will therefore make the ansatz
$N_\text{DC}=N_{\text{DC}}^{(0)}+(C_p/2C_S)N_{\text{DC}}^{(1)}+(C_p/2C_S)^2N_{\text{DC}}^{(2)}$
and we also expand the DC drive $v=v_0+(C_p/2C_S)v_1+(C_p/2C_S)^2v_2$ in the
same way. The result for $N_{\text{DC}}^{(0)}$ can be found in
Eq.~\eqref{eq:solutionDC}. In order to obtain the higher order contributions,
we separate Eq.~\eqref{eq:lintwice} and arrive at the expression
\begin{align}\label{eq:separation}
\int dN_\text{DC}\bigg\lbrace &v_0-\sin(N_\text{DC})
+(C_p/2C_S)\left[v_1+\sin(2N_\text{DC})\right]\nonumber\\
&+(C_p/2C_S)^2\left[v_2-\sin(N_\text{DC})-\sin(3N_\text{DC})\right] \bigg\rbrace ^{-1}
=\omega_R (t-t_0).
\end{align}
Now, we expand the integrand up to second order in small $C_p/C_S$ and perform
the resulting integrals. Then we can insert the expansion for $N_{DC}$ and
sort the expression in orders of $C_p/C_S$. While the zeroth order term will
lead to Eq.~\eqref{eq:solutionDC}, solving the higher order terms in
succession results in analytical expressions for $N_{\text{DC}}^{(1)}$ and
$N_{\text{DC}}^{(2)}$. Since both expressions are quite involved, we will
refrain from writing them down here. As a next step we will choose $v_1$ and $v_2$
in such a way that $\overline{\dot N_{\text{DC}}^{(1)}},\,\overline{\dot
N_{\text{DC}}^{(2)}}=0$. This way the average current will only depend on the
zeroth order contribution and our former relation $i_0=I_0/V_c
R=\sqrt{v_0^2-1}$ will thus remain valid throughout the calculation. If we
apply the condition to our analytical results we obtain
\begin{align}\label{eq:fullstepposition}
&v_1(v_0)=0,\\
&v_2(v_0)=6v_0-4v_0^3+4(v_0^2-1)^{3/2}.
\end{align}
The first step at $I_0 = e \omega_0/\pi$ , thus, appears close to value
$V_0=V_c[v_0+(C_p/2C_S)^2v_2(v_0)]$ with $v_0=\sqrt{1+\omega_0^2/\omega_R^2}$.
In order to calculate the step height we need to use
Eq.~\eqref{eq:ConditionStep} again, which remains valid in the case $C_p\neq
0$. Going through the same steps as before, we finally obtain the height of the
first step
\begin{equation}\label{eq:FinalStep}
\Delta V=\frac{V_\text{ac}}{v_0}\left\lbrace1+\left(\frac{C_p}{2C_S}\right)^2\left[1+\frac{29}{96v_0^2}+\mathcal{O}\left(\frac{1}{v_0^3}\right)\right]+\mathcal{O}\left(\frac{C_p^3}{C_S^3}\right)\right\rbrace,
\end{equation}
valid to leading order in $V_\mathrm{ac}/V_c$
\subsection{Ground state approximation}

In the regime $R\gg R_Q$, the system can be described by the Hamiltonian in
Eq.~\eqref{eq:Ham_Qq} coupled to the equation of motion in
Eq.~\eqref{eq:Motion_Qc}. Experimentally, the regime $Z\simeq R_Q$ is of most
interest and our calculation will be done in the overdamped regime with
$\omega_c\gg\omega_R$. In order to obtain an analytical result, we will assume
the quantum plasma frequency to be large enough such that the relations
$\omega_q\gg\omega_0$ and $\omega_q\gg eV_c/\hbar$ are fulfilled. In this
case, the last two terms in \eqref{eq:Ham_Qq} can be neglected and the system
will remain in the ground state of the harmonic oscillator with the ground
state wave function given by Eq.~\eqref{eq:gs_wf}. Next, we need to
approximate the expectation value of $\hat{Q}_q/C_p$. Since we assume $C_p$ to
be a small parameter in our expansion we can not simply take the expectation
value of $\hat{Q}_q$ in the ground state. Instead we express $\hat{Q}_q/C_p$
in terms of the commutator $[\hat{H},\hat{\Phi}_q]$
\begin{align}\label{eq:Commutator}
\hat{Q}_q/C_p=[\hat{H},\hat{\Phi}_q]/i\hbar-L\ddot{Q}_c-V_c\sin(\pi[\hat{Q}_q+Q_c]/e).
\end{align}
Within our approximation, the expectation value of the commutator is zero and
we therefore only have to calculate the expectation value of
\begin{align}\label{eq:V_cSin}
\langle \psi_0|\sin(\pi[\hat{Q}_q+Q_c]/e)| \psi_0\rangle=e^{-\pi R_Q/2Z}\sin (\pi Q_c/e).
\end{align}
If we insert both results in Eq.~\eqref{eq:Motion_Qc}, we obtain a simplified
equation of motion for the classical charge, which can be found in
Eq.~\eqref{eq:Motion_Qc_Ground}.

\subsection{Influence of thermal noise}
Throughout the paper, we assume that the temperature is sufficiently low to keep the dual Shapiro steps from being washed out. Here, we want to comment on the influence of noise on the step height. 
First, we will discuss the influence of charge fluctuations in the $LC$-circuit composed of the parasitic capacitance $C_p$ and the superinductance $L$. If we include thermal excitation of higher energy levels in our ground state approximation, Eq.~\eqref{eq:V_cSin} has to be rewritten in terms of the trace with the density operator $\hat \rho$
\begin{align}\label{eq:Charge_noise}
\mathrm{Tr}\{\hat \rho\sin(\pi[\hat{Q}_q+Q_c]/e)\}=e^{-\pi R_Q(1+2\bar{n})/2Z}\sin (\pi Q_c/e), \qquad \bar{n}=1/(e^{\hbar \omega_q/k_B T}-1)
\end{align}
where $\bar{n}$ is the average number of photons in the $LC$-resonator given by the Bose-Einstein statistic. For a realistic resonance frequency of the order of $\omega_q \simeq 5\,\mathrm{Ghz}$ and a temperature of $T\simeq 20\,\mathrm{mK}$, we obtain $\bar{n}\simeq 0.1$. The influence of charge fluctuations in the $LC$-resonator is thus negligible.

Next, we will discuss the influence of the thermal noise of the resistance. There have been extensive studies on the influence of thermal, white noise for the RSJ-model of the conventional Shapiro step experiment \cite{likharev}, which can be easily transferred to the problem of dual Shapiro steps. As an effective noise parameter, we obtain
\begin{align}\label{eq:Charge_noise}
\gamma=2\pi\frac{k_B T}{e \Delta V}\frac{R}{R_d(V_0)}\left(1+\frac{V_c^2}{2V_0^2}\right),
\end{align}
with the differential resistance $R_d(V_0)=RV_c\sqrt{(V_0/V_c)^2-1}/V_0$ and $V_0$ the DC-voltage around which the first dual Shapiro step is centered. The washing out of the step takes place at $\gamma \sim 1$, which leads to a suppression of steps smaller than $\Delta V\simeq V_c$ ($V_c\simeq 10 \, \mu \mathrm{V}$) at a temperature of $T\simeq 20\,\mathrm{mK}$. This would suggest that for a resistor at $T\simeq 4\,$K even the measurement of Coulomb blockade is not possible. However, effective filtering methods can be employed to significantly lower the thermal noise of resistors at higher temperatures. This has been repeatedly demonstrated in Coulomb blockade measurements \cite{kuzmin:91,haviland:91,haviland:94,Corlevi:06}.


\begin{thebibliography}{10}
\makeatletter

\bibitem{likharev:85}
K. K. Likharev and A. B. Zorin,
 \href{http://dx.doi.org/10.1007/BF00683782}{%
 J. Low Temp. Phys. {\bf 59}, 347 (1985)}.

\bibitem{Pekola:13}
J. P. Pekola {\em et~al.\/},
 \href{http://dx.doi.org/10.1103/RevModPhys.85.1421}{%
 Rev. Mod. Phys. {\bf 85}, 1421 (2013)}.

\bibitem{Averin:85}
D. Averin, A. Zorin, and K. Likharev,
 Sov. Phys. JETP {\bf 61}, 407 (1985).

\bibitem{schmid:83}
A. Schmid,
 \href{http://dx.doi.org/doi.org/10.1103/PhysRevLett.51.1506}{%
 Phys. Rev. Lett. {\bf 51}, 1506 (1983)}.

\bibitem{grabert}
H. Grabert and M. H. Devoret, eds.,
 {\em Single Charge Tunneling\/}, vol. 294 of {\em NATO ASI Series B\/}
 (Plenum Press, New York, 1992).

\bibitem{mooij:06}
J. E. Mooij and {\relax Yu}. V. Nazarov,
 \href{http://dx.doi.org/10.1038/nphys234}{%
 Nature Phys. {\bf 2}, 169 (2006)}.

\bibitem{pop:10}
I. M. Pop {\em et~al.\/},
 \href{http://dx.doi.org/10.1038/nphys1697}{%
 Nature Phys. {\bf 6}, 589 (2010)}.

\bibitem{astafiev:12}
O. V. Astafiev {\em et~al.\/},
 \href{http://dx.doi.org/10.1038/nature10930}{%
 Nature {\bf 484}, 355 (2012)}.

\bibitem{masluk:12}
N. A. Masluk {\em et~al.\/},
 \href{http://dx.doi.org/10.1103/PhysRevLett.109.137002}{%
 Phys. Rev. Lett. {\bf 109}, 137002 (2012)}.

\bibitem{hongisto:12}
T. T. Hongisto and A. B. Zorin,
 \href{http://dx.doi.org/10.1103/PhysRevLett.108.097001}{%
 Phys. Rev. Lett. {\bf 108}, 097001 (2012)}.

\bibitem{Lehtinen:12}
J. S. Lehtinen, K. Zakharov, and K. Y. Arutyunov,
 \href{http://dx.doi.org/10.1103/PhysRevLett.109.187001}{%
 Phys. Rev. Lett. {\bf 109}, 187001 (2012)}.

\bibitem{haviland:94}
D. B. Haviland {\em et~al.\/},
 \href{http://dx.doi.org/10.1103/PhysRevLett.73.1541}{%
 Phys. Rev. Lett. {\bf 73}, 1541 (1994)}.

\bibitem{Corlevi:06}
S. Corlevi, W. Guichard, F. W. J. Hekking, and D. B. Haviland,
 \href{http://dx.doi.org/10.1103/PhysRevB.74.224505}{%
 Phys. Rev. B {\bf 74}, 224505 (2006)}.

\bibitem{kuzmin:91}
L. S. Kuzmin and D. B. Haviland,
 \href{http://dx.doi.org/10.1103/PhysRevLett.67.2890}{%
 Phys. Rev. Lett. {\bf 67}, 2890 (1991)}.

\bibitem{haviland:91}
D. B. Haviland, L. S. Kuzmin, P. Delsing, and T. Claeson,
 EPL {\bf 16} (1), 103 (1991).

\bibitem{manucharyan}
V. E. Manucharyan,
 PhD thesis, Yale University, 2012,
 chapter~5.4.

\bibitem{koch:09}
J. Koch, V. Manucharyan, M. H. Devoret, and L. I. Glazman,
 \href{http://dx.doi.org/10.1103/PhysRevLett.103.217004}{%
 Phys. Rev. Lett. {\bf 103}, 217004 (2009)}.

\bibitem{guichard:10}
W. {Guichard} and F. W. J. {Hekking},
 \href{http://dx.doi.org/10.1103/PhysRevB.81.064508}{%
 \prb {\bf 81} (6), 064508 (2010)}.

\bibitem{marco:15}
A. D. Marco, F. W. J. Hekking, and G. Rastelli,
 \href{http://dx.doi.org/10.1103/PhysRevB.91.184512}{%
 Phys. Rev. B {\bf 91}, 184512 (2015)}.

\bibitem{manucharyan:09}
V. E. Manucharyan, J. Koch, L. Glazman, and M. Devoret,
 \href{http://dx.doi.org/10.1126/science.1175552}{%
 Science {\bf 326}, 113 (2009)}.

\bibitem{bell:12}
M. T. Bell {\em et~al.\/},
 \href{http://dx.doi.org/10.1103/PhysRevLett.109.137003}{%
 Phys. Rev. Lett. {\bf 109}, 137003 (2012)}.

\bibitem{altimiras:13}
C. Altimiras {\em et~al.\/},
 \href{http://dx.doi.org/10.1063/1.4832074}{%
 Appl. Phys. Lett. {\bf 103}, 212601 (2013)}.

\bibitem{weissl:15}
T. Wei\ss{}l {\em et~al.\/},
 \href{http://dx.doi.org/10.1103/PhysRevB.91.014507}{%
 Phys. Rev. B {\bf 91}, 014507 (2015)}.

\bibitem{Fink:16}
J. M. Fink {\em et~al.\/},
 \href{http://dx.doi.org/10.1038/ncomms12396}{%
 Nature Commun. {\bf 7}, 12396 (2016)}.

\bibitem{zorin:06}
A. B. Zorin,
 \href{http://dx.doi.org/10.1103/PhysRevLett.96.167001}{%
 Phys. Rev. Lett. {\bf 96}, 167001 (2006)}.

\bibitem{houck:09}
A. A. Houck {\em et~al.\/},
 \href{http://dx.doi.org/10.1007/s11128-009-0100-6}{%
 Quant. Inf. Proc. {\bf 8}, 105 (2009)}.

\bibitem{koch:07}
J. Koch {\em et~al.\/},
 \href{http://dx.doi.org/10.1103/PhysRevA.76.042319}{%
 Phys. Rev. A {\bf 76}, 042319 (2007)}.

\bibitem{ulrich:16}
J. Ulrich and F. Hassler,
 \href{http://dx.doi.org/10.1103/PhysRevB.94.094505}{%
 Phys. Rev. B {\bf 94}, 094505 (2016)}.

\bibitem{suppl}
See Supplementary Material.

\bibitem{commut}
Note that in the loop charge description, charge acts as position
  \cite{ulrich:16}.

\bibitem{likharev}
K. K. Likharev,
 {\em Dynamics of {J}osephson Junctions and Circuits\/}
 (Gordon and Breach Science Publishers, 1986).

\end{thebibliography}
\end{document}